\documentclass[prl,showpacs,superscriptaddress,twocolumn,floatfix]{revtex4-2}
\usepackage{graphicx,dcolumn}
\usepackage{color}
\newcommand{\pllptte}{P_{LL} -P_{TT} / \epsilon}
\newcommand{\plt}{P_{LT}}
\newcommand{\ptt}{P_{TT}}
\newcommand{\qcm}{q_\mathrm{cm}}
\newcommand{\qpr}{q'_\mathrm{cm}}
\newcommand{\thcm}{\theta_{\mathrm{cm}}}
\newcommand{\phicm}{\varphi}
\newcommand{\cthcm}{\cos \theta_{\mathrm{cm}}}
\newcommand{\epg}{e p \to e p \gamma}

\newcommand{\aeq}{\alpha_{E1}(Q^2)}
\newcommand{\bmq}{\beta_{M1}(Q^2)}
\newcommand{\ale}{\alpha_{E1}}
\newcommand{\bem}{\beta_{M1}}
\newcommand{\ohigher}{{\cal O}(q_\mathrm{cm}^{\prime 2})}
\newcommand{\ohigherdr}{{\cal O}(q_\mathrm{cm}^{\prime 2})_{DR}}
\DeclareMathAccent{\pol}{\mathord}{letters}{"7E}
\begin{document}
\title{
New Insight in the $Q^2$-Dependence of Proton Generalized Polarizabilities}
%
%
\def\kph{\affiliation{Institut f\"ur Kernphysik, Johannes  Gutenberg-Universit\"at Mainz, D-55099 Mainz, Germany}}
\def\clermont{\affiliation{Universit\'e Clermont Auvergne, CNRS/IN2P3, LPC, F-63000 Clermont-Ferrand, France}}
\def\zagreb{\affiliation{Department of Physics, Faculty of Science, University of Zagreb, 10000   Zagreb, Croatia}}
\def\stefan{\affiliation{Jo\v zef Stefan Institute, SI-1000 Ljubljana, Slovenia}}
\def\unil{\affiliation{Faculty of Mathematics and Physics, University of Ljubljana, SI-1000 Ljubljana,  Slovenia}}  
\def\uniwash{\affiliation{Institute  for  Nuclear  Studies,  Department  of  Physics, The  George  Washington  University,  Washington  DC  20052,  USA}}
\def\utemple{\affiliation{Temple University, Philadelphia, PA 19122, USA}}
\def\mitlns{\affiliation{Laboratory for Nuclear Science, Massachussetts Institute of Technology, Cambridge, MA 02139, USA}}
\def\stonybrook{\affiliation{Department of Physics and Astronomy, Stony Brook University, SUNY, Stony Brook, NY 11794-3800, USA}}
\def\rikenbnl{\affiliation{RIKEN BNL Research Center, Brookhaven National Laboratory, Upton, NY 11973-5000, USA}}
%
\author{J.~Beri\v{c}i\v{c}}\stefan
\author{L.~Correa}\clermont\kph  
\author{M.~Benali}\clermont
\author{P.~Achenbach}\kph      
\author{C.~Ayerbe~Gayoso}\altaffiliation{Now at College of William and Mary, Williamsburg, VA 23185, USA}\kph 
\author{J.C.~Bernauer}\stonybrook\rikenbnl 
\author{A.~Blomberg}\utemple 
\author{R.~B\"ohm}\kph
\author{D.~Bosnar}\zagreb      
\author{L.~Debenjak}\stefan  
\author{A.~Denig}\kph
\author{M.O.~Distler}\kph
\author{E.J.~Downie}\uniwash        
\author{A.~Esser}\kph           
\author{H.~Fonvieille}\email[]{helene.fonvieille@clermont.in2p3.fr}\clermont
\author{I.~Fri\v{s}\v{c}i\'{c}}\mitlns
\author{S.~Kegel}\kph
\author{Y.~Kohl}\kph
\author{M.~Makek}\zagreb       
\author{H.~Merkel}\kph
\author{D.G.~Middleton}\kph
\author{M.~Mihovilovi\v{c}}\kph\stefan
\author{U.~M\"uller}\kph
\author{L.~Nungesser}\kph
\author{M.~Paolone}\utemple 
\author{J.~Pochodzalla}\kph         
\author{S.~S\'anchez Majos}\kph
\author{B.S.~Schlimme}\kph
\author{M.~Schoth}\kph
\author{F.~Schulz}\kph
\author{C.~Sfienti}\kph 
\author{S.~\v{S}irca}\unil\stefan
\author{N.~Sparveris}\utemple 
\author{S.~\v{S}tajner}\stefan
\author{M.~Thiel}\kph
\author{A.~Tyukin}\kph
\author{A.~Weber}\kph       
\author{M.~Weinriefer}\kph
\collaboration{A1 Collaboration}
\date{\today}
\begin{abstract}
 Virtual Compton scattering on the proton has been investigated at three yet unexplored values of the four-momentum transfer $Q^2$: 0.10, 0.20 and 0.45 GeV$^2$, at the  Mainz Microtron. Fits performed using either the low-energy theorem or dispersion relations allowed the extraction of the structure functions $\pllptte$ and $\plt$, as well as the electric and magnetic generalized polarizabilities $\aeq$ and $\bmq$. These new results show a smooth and rapid fall-off of $\aeq$, in contrast to previous measurements at $Q^2$ = 0.33 GeV$^2$, and provide for the first time a precise mapping of $\bmq$ in the low-$Q^2$ region.
\end{abstract}
\maketitle



The Compton scattering process gives access to one of the fundamental characteristics of the nucleon: its polarizabilities, i.e., the way the particle deforms under the influence of a quasi-static external electric or magnetic field. Real Compton scattering (RCS) on the proton, $\gamma p \to \gamma p$, has provided essential knowledge on the electric and magnetic polarizabilities $\ale$ and $\bem$. Virtual Compton scattering (VCS) has been shown to give access to new observables called generalized polarizabilities (GPs)~\cite{Arenhoevel:1974twc}.  The GPs generalize the polarizabilities of RCS to non-zero $Q^2$, and as such they yield the spatial distribution of polarizability via a Fourier transform~\cite{Gorchtein:2009qq}. The formalism of VCS for the nucleon case was first worked out in Ref.~\cite{Guichon:1995pu}. The nucleon polarizability phenomenon possesses several unique facets of interest, among which are a high sensitivity to the mesonic cloud and the interplay between dia- and paramagnetism.

VCS on the proton, $\gamma^* p \to \gamma p$, has been investigated at different laboratories (MAMI~\cite{Roche:2000ng,Bensafa:2006wr,Janssens:2008qe,Doria:2015dyx,Blomberg:2019caf}, MIT-Bates~\cite{Bourgeois:2006js,Bourgeois:2011zz}, JLab \cite{Laveissiere:2004nf,Fonvieille:2012cd}) and photon virtualities $Q^2$ between 0.06 GeV$^2$ and 1.76 GeV$^2$. A picture of the $Q^2$-dependence of the scalar GPs of the proton, i.e., the electric and  magnetic GPs $\aeq$ and $\bmq$, has slowly emerged from the measurements. However, due to the scarcity of the data and the difficulty of these experiments, our knowledge of the GPs is far from being complete or satisfactory. The main motivation of the present experiment was to provide new and precise data in order to build a more consistent picture of the electric and magnetic GPs of the proton.

VCS is measured through the $\epg$ reaction, for small values of the final real photon energy, $\qpr$,  in the center of mass (c.m.) of the virtual photon and initial proton ($\qpr \sim$ 100  MeV/$c$). In this regime the cross section can be decomposed into the coherent sum of the Bethe-Heitler, the Born and the non-Born processes, the latter being parametrized by the GPs. The formalism of Ref.~\cite{Guichon:1995pu} paved the way to extract the GPs via a low-energy theorem, or low-energy expansion (LEX). A few years later, the dispersion relation (DR) model has been developed for VCS~\cite{Pasquini:2000pk,Pasquini:2001yy,Drechsel:2002ar} and allowed for another method of experimental investigation of the GPs.

\section{The experiment}
\label{sec-the-experiment}

In the experiment we have used the MAMI accelerator beam, the 5 cm long liquid hydrogen target and the two high-resolution magnetic spectrometers A and B of the A1 setup~\cite{Blomqvist:1998xn}, to detect in coincidence the scattered electrons and recoil protons. Data were taken at three values of $Q^2$: 0.10, 0.20 and 0.45 GeV$^2$~\cite{BericicPhD:2015,CorreaPhD:2016,BenaliPhD:2016}, essentially below the pion production threshold ($\qpr \le$  126 MeV/$c$).
At each $Q^2$ the data were acquired in three types of kinematical settings: an in-plane setting with mixed sensitivity to $\aeq$ and $\bmq$, an out-of-plane setting with enhanced sensitivity to $\aeq$, and a low-$\qpr$ setting insensitive to GPs but useful for normalization purposes. Table \ref{tab-1-kinematics} gives a brief overview of the kinematics. By exploiting the out-of-plane capability of spectrometer B a large domain was covered in  $\thcm$ and $\phicm$, i.e., the polar and azimuthal angles of the real photon with respect to the virtual photon in the c.m. This ensured a large lever arm in the physics fits. The kinematics were also chosen to maximize the virtual photon polarization parameter  $\epsilon$ (cf. Table \ref{tab-results-lex-dr-fits}).

\begin{table}
\caption{\label{tab-1-kinematics}
The range covered by the kinematical settings, in beam energy, spectrometer momenta and angles, including the out-of-plane angle of spectrometer B ($OOP_B$). The scattered electron is detected in spectrometer B (resp. A) at the two lowest (resp. the highest) $Q^2$.}
\begin{ruledtabular}
  \begin{tabular}{ccccccc}
$Q^2$ & $E_{\mathrm{beam}}$ &  $p_A$  & $\theta_A$ &  $p_B$  & $\theta_B$ & $OOP_B$ \\
(GeV$^2$) & (MeV) & (MeV/$c$) & ($^{\circ}$) &  (MeV/$c$) & ($^{\circ}$) & ($^{\circ}$) \\
\noalign{\smallskip}
\hline
\noalign{\smallskip}
0.10 & 872 & 340-425 & 53-58 & 700-745  & 22-23 & 0-9.0 \\
0.20 & 905-1002 & 440-580 & 51-52 & 710-770 & 30-33 & 0-8.5 \\
0.45 & 937-1034 & 645-650 & 51-52 & 630-750 & 33-41 & 0-8.0 \\
\noalign{\smallskip}
%
  \end{tabular}
\end{ruledtabular}
\end{table}


The experiment was performed in several phases between 2011 and 2015. The unpolarized electron beam with beam current in the range 5--15 $\mu$A was sent to the target with a raster pattern.  The continuous monitoring of the target pressure and temperature ensured a stable liquid hydrogen density. The experimental luminosity was determined precisely using beam current measurement by a fluxgate magnetometer. The event rate was corrected for acquisition deadtime and a small scintillator inefficiency. The efficiency of the vertical drift chambers was considered to be  100\%.


An important step of the analysis is the calibration of experimental parameters. Spectrometer optics needed specific studies for settings where the magnets were in the saturation region, in order to achieve an optimal particle reconstruction at the vertex. Then the missing mass squared in $p(e,e'p)X$ was used as the main tool to optimize the various offsets  in momenta and angles. It was also used to determine the thickness of the cryogenic deposit on the target walls, which was an important parameter, especially at $Q^2=0.1$ GeV$^2$. After a careful calibration we estimate that the accuracy reached on these items guarantees the control of the solid angle to about  $\pm$ 1\%.

To extract a clean signal, a few main cuts were applied to the data. Firstly, true coincidences were selected by a cut around the narrow coincidence time peak with FWHM in the range 0.8--1.7 ns. Random coincidences were subtracted using side bands of the coincidence time spectrum. Secondly, a cut was applied in the longitudinal vertex coordinate in order to eliminate events coming from the target walls, mostly due to quasi-elastic $(e,e'p)$ reactions on nuclei. This reduced the useful target cell length to about $3 \, \mathrm{cm}$. Thirdly, the exclusive reaction $\epg$ was identified by the missing mass technique. Events were selected  in a window of [$-6 \sigma, +7 \sigma$] around the center of the photon peak in the missing mass squared spectrum, where $\sigma$ is the r.m.s. of the peak ($\sigma$ = 150 to 550 MeV$^2$ depending on kinematics). At this level the event sample was very clean and no particle-identification cuts were necessary.


A Monte-Carlo simulation of the experiment including all resolution effects and radiative corrections was used to determine the five-fold solid angle~\cite{Janssens:2006vx}. After having applied the analysis cuts to the simulated events, one obtained the absolute cross section $d^5 \sigma_\mathrm{exp} / ( d E_e' d \Omega_e' d \cthcm d \phicm ) $, denoted by $\sigma_\mathrm{exp}$ in the following. 
%
%
For each of the three $Q^2$, cross sections were determined at fixed virtual photon c.m. momentum  $\qcm$ and fixed $\epsilon$ (cf. Table \ref{tab-results-lex-dr-fits})  and variable ($\qpr, \cthcm, \phicm$). Due to the rapidly varying effect of the GPs in this three-dimensional phase space, a small bin size was chosen: 25 MeV/$c$ in $\qpr$, 0.05 in $\cthcm$ and 10$^{\circ}$ in $\phicm$. This resulted in many cross-section points, up to 10$^3$ per $Q^2$  (cf. the  n.d.f. column in Table \ref{tab-results-lex-dr-fits}).


Cross sections obtained for $\qpr \le$ 50 MeV/$c$ are used to test the normalization of the experiment. Indeed, at these very low photon energies the cross section is almost entirely given by the Bethe-Heitler+Born (BH+B) process, plus a very small GP effect ($\le$1\%). In these conditions, the comparison of the calculated cross section to the measured one provides unambiguously the renormalization factor to apply to $\sigma_\mathrm{exp}$. These factors are found to be close to 1 within $\pm$ [1-2]\%. For this test, as well as for the extraction of VCS observables, a choice of proton form factors is needed, namely to compute the BH+B cross section, $\sigma_\mathrm{BH+B}$. In the following, we use the parametrization  of $G_E^p$ and  $G_M^p$ of Ref.~\cite{Friedrich:2003iz}. It should be noted that the VCS physics results become practically independent of the choice of proton form factors, when the analysis is done consistently with the same form factor choice from the renormalization step to the physics fits; see for instance Refs.~\cite{BericicPhD:2015,CorreaPhD:2016,BenaliPhD:2016}.

\section{LEX and DR fits}
\label{sec-lex-and-dr-fits}

\begin{table*}
\caption{\label{tab-results-lex-dr-fits}
Results of the LEX and DR fits, all performed in the $\qpr$-range [50,125] MeV/$c$. The first error is statistical. The second one is the total systematic error, whose sign indicates the correlation to the ($\pm$) sign of the overall normalization change (see text). For each $Q^2$ and each type of fit (LEX or DR) the first line is obtained with $K = 0.025$ (see text) and the second line, containing numbers in parentheses, is obtained without bin exclusion. In the LEX part of the table, the GPs are obtained only indirectly, by subtracting from the structure functions the spin-GP contribution calculated by the DR model.}
\begin{ruledtabular}
  \begin{tabular}{cccccccc}
$Q^2$  & $\qcm$ & $\epsilon$  &  $\pllptte$  &  $\plt$ & $\aeq$  & $\bmq$ & reduced $\chi^2$ \\
(GeV$^2$) &  (MeV/$c$) & &  ($\mathrm{GeV}^{-2}$) &  ($\mathrm{GeV}^{-2}$) &  (10$^{-4}$fm$^3$) &  (10$^{-4}$fm$^3$) & / n.d.f. \\
%
\hline
\noalign{\smallskip}
\multicolumn{8}{c}{LEX fit}   \\
\noalign{\smallskip}
\hline
\noalign{\smallskip}
%
%
0.10 & 320 & 0.91 &   33.15 $\pm$ 1.53 $\mp$ 4.53  &  $-$8.54  $\pm$ 0.60    $\mp$ 1.62  &  6.06  $\pm$ 0.30   $\mp$ 0.90 &  2.82   $\pm$ 0.23   $\pm$ 0.63 &  1.30/460 \\
\ & \ & \ & (23.31 $\pm$ 0.92 $\mp$ 4.11)  &  ($-$7.89  $\pm$ 0.33    $\mp$ 1.56)  &  (4.11  $\pm$ 0.18   $\mp$ 0.81)  &  (2.57   $\pm$ 0.13   $\pm$ 0.61)  & (1.63/707)  \\
\noalign{\smallskip}
\hline
\noalign{\smallskip}
%
0.20 & 458 & 0.85 &  14.57  $\pm$ 0.55   $\mp$ 3.47  & $-$5.37  $\pm$ 0.33   $\mp$ 1.25  & 3.02   $\pm$ 0.14  $\mp$ 0.87  &  2.01  $\pm$ 0.16   $\pm$ 0.61 & 1.29/1034  \\
\ & \ & \ &  (12.13  $\pm$ 0.42   $\mp$ 3.16)  &  ($-$5.02  $\pm$ 0.21   $\mp$ 1.28)  &  (2.40   $\pm$ 0.10  $\mp$ 0.79)  &  (1.84  $\pm$ 0.10   $\pm$ 0.62)  & (1.49/1231) \\  
%
\noalign{\smallskip}
\hline
\noalign{\smallskip}
0.45 & 714 & 0.63 & \, 4.21   $\pm$ 0.65   $\mp$ 2.24 & $-$1.00  $\pm$ 0.37   $\mp$ 0.50 & 0.92  $\pm$ 0.26  $\mp$ 0.92 &   0.19  $\pm$ 0.28   $\pm$ 0.38  & 1.17/820 \\ 
\ & \ & \  &  \, (4.33  $\pm$ 0.65   $\mp$ 2.23)  &  ($-$1.11  $\pm$ 0.36   $\mp$ 0.50)  & (0.97  $\pm$ 0.26  $\mp$ 0.91)  & (0.28  $\pm$ 0.28   $\pm$ 0.38)  & (1.17/839) \\
\noalign{\smallskip}
\hline
\noalign{\smallskip}
\multicolumn{8}{c}{DR fit}   \\
\noalign{\smallskip}
\hline
\noalign{\smallskip}
0.10 & 320 & 0.91  &  35.95 $\pm$ 1.80  $\mp$ 5.21 & \, $-$9.03  $\pm$  0.98  $\mp$ 1.82  & 6.60 $\pm$ 0.36 $\mp$ 1.03 &  3.02  $\pm$ 0.38  $\pm$ 0.72 &  1.34/460 \\
\ & \ & \ & (34.72 $\pm$ 1.24  $\mp$ 4.95)  &  ($-$10.40  $\pm$  0.58  $\mp$ 1.77)  & (6.35 $\pm$ 0.24 $\mp$ 0.98)   &  (3.55  $\pm$ 0.23  $\pm$ 0.69)  & (1.35/707) \\
\noalign{\smallskip}
\hline
\noalign{\smallskip}
%
0.20 & 458 & 0.85 &  14.94 $\pm$ 0.60 $\mp$ 4.06 &  $-$5.31  $\pm$ 0.44 $\mp$ 1.40 &   3.11  $\pm$ 0.15  $\mp$ 1.02  &  1.98  $\pm$ 0.22  $\pm$ 0.68  & 1.31/1034  \\ 
\ & \ & \ & (14.78 $\pm$ 0.50 $\mp$ 3.79)  &  ($-$5.83  $\pm$ 0.34 $\mp$ 1.49)  &  (3.07  $\pm$ 0.13  $\mp$ 0.95)  &   (2.24  $\pm$ 0.17  $\pm$ 0.73)  &   (1.34/1231)  \\ 
%
\noalign{\smallskip}
\hline
\noalign{\smallskip}
0.45 & 714 & 0.63  &  \, 4.10  $\pm$ 0.62  $\mp$ 2.48 &  $-$1.36 $\pm$ 0.29  $\mp$ 0.40 & 0.87 $\pm$ 0.25  $\mp$ 1.01 & 0.47 $\pm$ 0.22  $\pm$ 0.30 & 1.14/820 \\
\ & \ & \ &  \, (4.14  $\pm$ 0.62  $\mp$ 2.48) &  ($-$1.39 $\pm$ 0.29  $\mp$ 0.40) &  (0.89 $\pm$ 0.25  $\mp$ 1.01)  &  (0.49 $\pm$ 0.22  $\pm$ 0.30)  &  (1.14/839)  \\ 
\noalign{\smallskip}
  \end{tabular}
\end{ruledtabular}
\end{table*}

We first recall the $\qpr$-expansion of the  ($\epg$) cross section according to the LEX:
%
\begin{eqnarray}
\begin{array}{lll}
\sigma_\mathrm{LEX} & =  & \sigma_\mathrm{BH+B} +  ( \Phi \qpr ) \cdot  \Psi_0   +  \ohigher , \\
 \Psi_0  & =  & V_1  \cdot ( \pllptte ) + V_2 \cdot \plt  ,
\label{lexformula}
\end{array}
\end{eqnarray}
%
where $\sigma_\mathrm{BH+B}$ contains no polarizability effect and represents typically 90\% or more of the cross section below the pion production threshold. Here $ \Phi \qpr, V_1, V_2$ are known kinematical factors, and  the VCS response functions are the structure functions $P_{LL} \propto \aeq , \plt \propto \bmq$ + spin GPs, and $\ptt  \propto$ spin GPs (see~\cite{Guichon:1998xv} for details). This formula  provides the analytical expression of the first-order polarizability term ($\Psi_0$) but does not give a clue about the importance of the higher-order term  $\ohigher$, which depends on GPs of all orders. The LEX fit consists in comparing the measured cross section to the expression of Eq. (\ref{lexformula}) in its truncated form, i.e., neglecting  the  $\ohigher$ term. However, in some cases, this truncation is not reliable enough. In the MIT-Bates experiment~\cite{Bourgeois:2011zz}, the LEX analysis of the in-plane data for $\plt$ was found to be unreliable due to the smallness of the $\Psi_0$ term with respect to the $\ohigher$ term.


In our experiment, angular regions in ($\cthcm, \phicm)$  were selected to avoid such a potential difficulty. To this aim, we used the DR model. Indeed, a unique advantage of the DR framework is to include all orders in $\qpr$. Therefore, subtracting from the DR cross section, $\sigma_\mathrm{DR}$, the (truncated) LEX cross section, $\sigma_\mathrm{LEX}$, is a way to isolate the $\ohigher$ term of the LEX. More precisely, at any phase-space point and for a given set of input GPs, one can calculate the quantity $\ohigherdr = (\sigma_\mathrm{DR} - \sigma_\mathrm{LEX})/\sigma_\mathrm{BH+B}$ which  provides a valuable estimator of the importance of the $\ohigher$ term. We studied the behavior of the LEX fit when including bins with increasing values of $\ohigherdr$. This is realized by setting a cut $\vert \ohigherdr \vert \le K$, where $K$ acts as a threshold for bin exclusion, or ``bin masking". For a very tight cut, $K$ = 0.005, few bins are retained in the fit. For a very loose cut, e.g., $K$ = 0.15 at  $Q^2$ = 0.20 GeV$^2$,  no bins are excluded. Details of this study, including a fine scan in $K$, will be presented elsewhere. We found that $K$ = 0.025 is optimal for a LEX fit with bin masking in our kinematics. In practice, the computation of $\ohigherdr$  depends on input values for the structure functions, therefore the whole procedure (bin masking + LEX fit) needs a few iterations.


The results of the LEX fit with and without bin exclusion are presented in the upper half of Table \ref{tab-results-lex-dr-fits}. The difference between these two types of LEX fits increases when $Q^2$ decreases. The largest difference is observed  at $Q^2$ = 0.10 GeV$^2$ for $\pllptte$ (33 and 23  GeV$^{-2}$) and $\aeq$ (6 and 4 $\times 10^{-4}$ fm$^3$). As an outcome of this study, we consider the LEX results with bin masking as the most reliable ones.


The DR fit consists in comparing the measured cross section to the one calculated by the model for all possible values of its free parameters, which are an unconstrained  part of the two scalar GPs. By minimizing a $\chi^2$ one then finds  $\aeq$ and $\bmq$, as well as the structure functions $\pllptte$ and $\plt$.  In principle, the bin masking described above is not necessary for the DR fit. We nevertheless performed the study, and the results are presented  in the lower half of Table \ref{tab-results-lex-dr-fits}. In some cases, a good stability between the results with and without bin exclusion is acquired for the DR fit compared to the LEX fit, namely  for $\pllptte$  and $\aeq$  at $Q^2$ = 0.10 GeV$^2$. This confirms that the DR calculation gives a good account of the ($\epg$) cross section over a wide phase space and deals well with its $\qpr$-dependence.


It should be noted that the LEX and DR fits, when both performed with $(K=0.025)$, are expected to agree well mutually, because this condition selects bins where the two fitting hypotheses ($\sigma_\mathrm{LEX}$ and $\sigma_\mathrm{DR}$) are very close to each other. In a sense, these two fits are not fully independent, and the LEX fit acquires a slight (DR-)model dependence. On the other hand, the LEX fit without bin masking is independent of any input from DR.


Statistical errors on the physics observables are provided for each fit by the size of the contour at $\chi^2_\mathrm{min}+1$ ($\chi^2$ non-reduced). For the systematic error, part of it disappears when using the normalization based on the low-$\qpr$ data. This is the case for errors common to all settings, related for instance to luminosity determination or radiative corrections. The final systematic error on the physics observables is estimated by changing the overall normalization of the measured cross section by $\pm$1.5\%. It is a quick and efficient way to include all remaining systematic errors, as shown in Ref.~\cite{CorreaPhD:2016}. Here the $\pm$1.5\% overall uncertainty is obtained as the quadratic sum of three contributions: $\pm$1\% due to the low-$\qpr$ normalization procedure (i.e., the intrinsic normalization uncertainty), $\pm$1\% due to the calibration and solid angle calculation, and $\pm$0.5\% due to other sources, such as radiative corrections or form-factor choices.
Exceptionally, when this method does not work, the systematic error is taken from the spread observed between several analyses involving different calibrations and offsets; this is the case for $\plt$ and $\bmq$ at  $Q^2$ = 0.45 GeV$^2$.

\begin{figure}
\centerline{\includegraphics[width=\columnwidth]{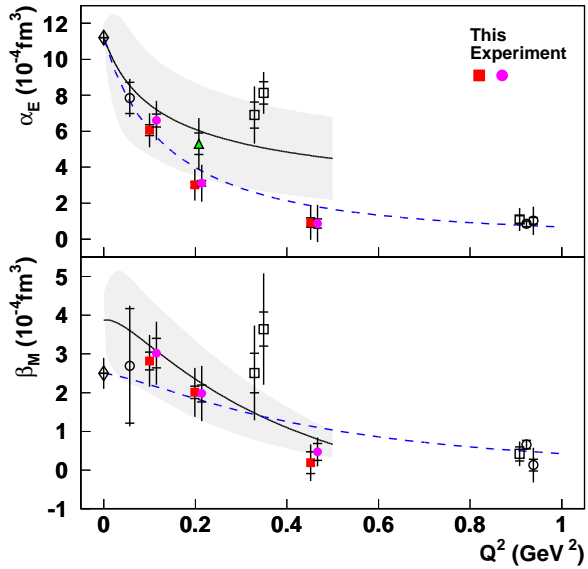}}
\caption{(Color online) The electric and magnetic GPs of the proton. Filled circles and squares at $Q^2$ = 0.10, 0.20 and 0.45 GeV$^2$  are from this experiment. Open circles and squares are from previous experiments at MIT-Bates~\cite{Bourgeois:2011zz}, MAMI~\cite{Roche:2000ng,Janssens:2008qe} and JLab~\cite{Fonvieille:2012cd}. Open and filled circles correspond to DR analyses; open and filled squares correspond to LEX analyses. The triangle point in $\aeq$ is from the recent experiment of Ref.~\cite{Blomberg:2019caf} and the RCS point ($\diamond$) is from Ref.~\cite{Tanabashi:2018oca}. The dashed curve is obtained using the DR model~\cite{Drechsel:2002ar} with dipole mass parameters $\Lambda_{\alpha} = \Lambda_{\beta}$ = 0.7 GeV. The dotted curve with its error band (shaded area) is from covariant B$\chi$PT~\cite{Lensky:2016nui}. Some data points are slightly shifted in abscissa for visibility. The inner and outer error bars are statistical and total, respectively.
}
\label{fig:1}
\end{figure}

Figure \ref{fig:1} displays the results of the experiment for the GPs together with the existing data. The new measurements show two clear features: the fall-off with $Q^2$ is both smooth and rapid. Our points connect smoothly to existing data, except to the (two independent) former measurements at $Q^2$ = 0.33 GeV$^2$~\cite{Roche:2000ng,Janssens:2008qe}  which lie above the general trend. Somewhat in the spirit of our bin masking, which constrains the cross section to follow a well-defined  $\qpr$-behavior, several fits were performed by the authors of the first VCS experiment~\cite{Roche:2000ng} on their data at $Q^2$ = 0.33 GeV$^2$, assuming different $\qpr$-evolutions of the cross section~\cite{dHose:2006bos}. All the fits of~\cite{dHose:2006bos}, including the DR one, lead to an enhanced value for $\pllptte$ [and hence $\aeq$], thus leaving the discrepancy unexplained for the electric GP. At the same time, the DR fit of~\cite{dHose:2006bos} brings $\plt$ [and hence $\bmq$] closer to zero by one standard deviation, making the magnetic GP agree with the general trend (this point is not shown in the figure). Our data point for the electric GP at $Q^2$ = 0.20 GeV$^2$ agrees with  the recent measurement of Ref.~\cite{Blomberg:2019caf} at the same $Q^2$, within experimental uncertainties. Taken together, these two new sets of data do not suggest any large enhancement of  $\aeq$, but leave room for a milder one. A forthcoming VCS experiment at JLab~\cite{Sparveris:2016} will provide more data at $Q^2$ = 0.33 GeV$^2$ and will shed light on this anomaly. Our measurements indicate a rapid decrease of the GPs with four-momentum transfer, with values at $Q^2$ = 0.45 GeV$^2$ being as small as the JLab ones at $Q^2$ = 0.92 GeV$^2$~\cite{Fonvieille:2012cd}. \newline
Our data at the smallest $Q^2$ connect nicely to the MIT-Bates VCS and the RCS points, in particular confirming the large mean square polarizability radius of $\ale$ of $\approx$ 2 fm$^2$~\cite{Bourgeois:2006js} which evinces mesonic cloud effects. The present experiment also provides the first precise measurement of the magnetic GP at low $Q^2$ (0.10 GeV$^2$), imposing a strong constraint on the  possible  extremum of $\bmq$ and on the understanding of its competing dia- and paramagnetic components. However, the precise value of $\bem (Q^2=0)$ is still under debate (see, e.g., Ref.~\cite{Pasquini:2019nnx}) and no meaningful mean square polarizability radius of $\bem$ can be quoted yet. The DR curve in Fig.~\ref{fig:1} accomodates well the experimental data for $\aeq$, which behave almost like a pure dipole. The baryon chiral perturbation theory (B$\chi$PT) calculation~\cite{Lensky:2016nui} accomodates  well  $\bmq$, although the theoretical uncertainty is large. As a final remark, measuring nucleon GPs is a challenging task due to the smallness of the polarizability effect. In our data sample the GP effect reaches at most 15\% of the ($\epg$) cross section at $Q^2$ = 0.10 and  0.20 GeV$^2$, and only 5\%  at $Q^2$ = 0.45 GeV$^2$ due to a lower $\epsilon$.

In summary, a high-statistics VCS experiment performed at MAMI has yielded precise measurements of the proton electric and magnetic GPs at three yet unexplored values of $Q^2$. Absolute $(\epg)$ cross sections have been measured below the pion production threshold over a wide range in phase space. The reliability of the LEX fit was improved by a novel treatment of the higher-order terms of the expansion. The dispersion relation model has proven once again its appropriateness for analyzing VCS at low energy. These new data indicate a smooth  $Q^2$-behavior  of the GPs, in contrast to some of the previous measurements. They help  building a coherent picture of the GPs and will strongly constrain future model calculations of nucleon GPs.


\vskip 2 mm
We would like to thank Barbara Pasquini, Marc Vanderhaeghen, Vladimir Pascalutsa and Vadim Lensky for their theoretical support. We gratefully acknowledge the MAMI-C accelerator group for the excellent beam quality. This work was supported by the Deutsche For\-schungsgemeinschaft with the Collaborative Research Center 1044, the Federal State of Rhineland-Palatinate and the French CNRS/IN2P3.

\bibliographystyle{apsrev.bst} 
\bibliography{vcsbiblio}

\begin{thebibliography}{27}
\expandafter\ifx\csname natexlab\endcsname\relax\def\natexlab#1{#1}\fi
\expandafter\ifx\csname bibnamefont\endcsname\relax
  \def\bibnamefont#1{#1}\fi
\expandafter\ifx\csname bibfnamefont\endcsname\relax
  \def\bibfnamefont#1{#1}\fi
\expandafter\ifx\csname citenamefont\endcsname\relax
  \def\citenamefont#1{#1}\fi
\expandafter\ifx\csname url\endcsname\relax
  \def\url#1{\texttt{#1}}\fi
\expandafter\ifx\csname urlprefix\endcsname\relax\def\urlprefix{URL }\fi
\providecommand{\bibinfo}[2]{#2}
\providecommand{\eprint}[2][]{\url{#2}}

\bibitem[{\citenamefont{Arenh{\"o}vel and Drechsel}(1974)}]{Arenhoevel:1974twc}
\bibinfo{author}{\bibfnamefont{H.}~\bibnamefont{Arenh{\"o}vel}}
  \bibnamefont{and} \bibinfo{author}{\bibfnamefont{D.}~\bibnamefont{Drechsel}},
  \bibinfo{journal}{Nucl. Phys.} \textbf{\bibinfo{volume}{A233}},
  \bibinfo{pages}{153} (\bibinfo{year}{1974}).

\bibitem[{\citenamefont{Gorchtein et~al.}(2010)\citenamefont{Gorchtein, Lorce,
  Pasquini, and Vanderhaeghen}}]{Gorchtein:2009qq}
\bibinfo{author}{\bibfnamefont{M.}~\bibnamefont{Gorchtein}},
  \bibinfo{author}{\bibfnamefont{C.}~\bibnamefont{Lorce}},
  \bibinfo{author}{\bibfnamefont{B.}~\bibnamefont{Pasquini}}, \bibnamefont{and}
  \bibinfo{author}{\bibfnamefont{M.}~\bibnamefont{Vanderhaeghen}},
  \bibinfo{journal}{Phys. Rev. Lett.} \textbf{\bibinfo{volume}{104}},
  \bibinfo{pages}{112001} (\bibinfo{year}{2010}), \eprint{0911.2882}.

\bibitem[{\citenamefont{Guichon et~al.}(1995)\citenamefont{Guichon, Liu, and
  Thomas}}]{Guichon:1995pu}
\bibinfo{author}{\bibfnamefont{P.~A.~M.} \bibnamefont{Guichon}},
  \bibinfo{author}{\bibfnamefont{G.~Q.} \bibnamefont{Liu}}, \bibnamefont{and}
  \bibinfo{author}{\bibfnamefont{A.~W.} \bibnamefont{Thomas}},
  \bibinfo{journal}{Nucl. Phys.} \textbf{\bibinfo{volume}{A591}},
  \bibinfo{pages}{606} (\bibinfo{year}{1995}), \eprint{nucl-th/9605031}.

\bibitem[{\citenamefont{Roche et~al.}(2000)}]{Roche:2000ng}
\bibinfo{author}{\bibfnamefont{J.}~\bibnamefont{Roche}} \bibnamefont{et~al.}
  (\bibinfo{collaboration}{A1}), \bibinfo{journal}{Phys. Rev. Lett.}
  \textbf{\bibinfo{volume}{85}}, \bibinfo{pages}{708} (\bibinfo{year}{2000}),
  \eprint{hep-ex/0007053}.

\bibitem[{\citenamefont{Bensafa et~al.}(2007)}]{Bensafa:2006wr}
\bibinfo{author}{\bibfnamefont{I.~K.} \bibnamefont{Bensafa}}
  \bibnamefont{et~al.} (\bibinfo{collaboration}{A1}), \bibinfo{journal}{Eur.
  Phys. J.} \textbf{\bibinfo{volume}{A32}}, \bibinfo{pages}{69}
  (\bibinfo{year}{2007}), \eprint{hep-ph/0612248}.

\bibitem[{\citenamefont{Janssens et~al.}(2008)}]{Janssens:2008qe}
\bibinfo{author}{\bibfnamefont{P.}~\bibnamefont{Janssens}} \bibnamefont{et~al.}
  (\bibinfo{collaboration}{A1}), \bibinfo{journal}{Eur. Phys. J.}
  \textbf{\bibinfo{volume}{A37}}, \bibinfo{pages}{1} (\bibinfo{year}{2008}),
  \eprint{0803.0911}.

\bibitem[{\citenamefont{Doria et~al.}(2015)}]{Doria:2015dyx}
\bibinfo{author}{\bibfnamefont{L.}~\bibnamefont{Doria}} \bibnamefont{et~al.}
  (\bibinfo{collaboration}{A1}), \bibinfo{journal}{Phys. Rev.}
  \textbf{\bibinfo{volume}{C92}}, \bibinfo{pages}{054307}
  (\bibinfo{year}{2015}), \eprint{1505.06106}.

\bibitem[{\citenamefont{Blomberg et~al.}(2019)}]{Blomberg:2019caf}
\bibinfo{author}{\bibfnamefont{A.}~\bibnamefont{Blomberg}} \bibnamefont{et~al.}
  (\bibinfo{year}{2019}), \bibinfo{note}{to appear in EPJA},
  \eprint{1901.08951}.

\bibitem[{\citenamefont{Bourgeois et~al.}(2006)}]{Bourgeois:2006js}
\bibinfo{author}{\bibfnamefont{P.}~\bibnamefont{Bourgeois}}
  \bibnamefont{et~al.}, \bibinfo{journal}{Phys. Rev. Lett.}
  \textbf{\bibinfo{volume}{97}}, \bibinfo{pages}{212001}
  (\bibinfo{year}{2006}), \eprint{nucl-ex/0605009}.

\bibitem[{\citenamefont{Bourgeois et~al.}(2011)}]{Bourgeois:2011zz}
\bibinfo{author}{\bibfnamefont{P.}~\bibnamefont{Bourgeois}}
  \bibnamefont{et~al.}, \bibinfo{journal}{Phys. Rev.}
  \textbf{\bibinfo{volume}{C84}}, \bibinfo{pages}{035206}
  (\bibinfo{year}{2011}).

\bibitem[{\citenamefont{Laveissiere et~al.}(2004)}]{Laveissiere:2004nf}
\bibinfo{author}{\bibfnamefont{G.}~\bibnamefont{Laveissiere}}
  \bibnamefont{et~al.} (\bibinfo{collaboration}{Jefferson Lab Hall A}),
  \bibinfo{journal}{Phys. Rev. Lett.} \textbf{\bibinfo{volume}{93}},
  \bibinfo{pages}{122001} (\bibinfo{year}{2004}), \eprint{hep-ph/0404243}.

\bibitem[{\citenamefont{Fonvieille et~al.}(2012)}]{Fonvieille:2012cd}
\bibinfo{author}{\bibfnamefont{H.}~\bibnamefont{Fonvieille}}
  \bibnamefont{et~al.} (\bibinfo{collaboration}{Jefferson Lab Hall A}),
  \bibinfo{journal}{Phys.Rev.} \textbf{\bibinfo{volume}{C86}},
  \bibinfo{pages}{015210} (\bibinfo{year}{2012}), \eprint{1205.3387}.

\bibitem[{\citenamefont{Pasquini et~al.}(2000)\citenamefont{Pasquini, Drechsel,
  Gorchtein, Metz, and Vanderhaeghen}}]{Pasquini:2000pk}
\bibinfo{author}{\bibfnamefont{B.}~\bibnamefont{Pasquini}},
  \bibinfo{author}{\bibfnamefont{D.}~\bibnamefont{Drechsel}},
  \bibinfo{author}{\bibfnamefont{M.}~\bibnamefont{Gorchtein}},
  \bibinfo{author}{\bibfnamefont{A.}~\bibnamefont{Metz}}, \bibnamefont{and}
  \bibinfo{author}{\bibfnamefont{M.}~\bibnamefont{Vanderhaeghen}},
  \bibinfo{journal}{Phys. Rev.} \textbf{\bibinfo{volume}{C62}},
  \bibinfo{pages}{052201} (\bibinfo{year}{2000}), \eprint{hep-ph/0007144}.

\bibitem[{\citenamefont{Pasquini et~al.}(2001)\citenamefont{Pasquini,
  Gorchtein, Drechsel, Metz, and Vanderhaeghen}}]{Pasquini:2001yy}
\bibinfo{author}{\bibfnamefont{B.}~\bibnamefont{Pasquini}},
  \bibinfo{author}{\bibfnamefont{M.}~\bibnamefont{Gorchtein}},
  \bibinfo{author}{\bibfnamefont{D.}~\bibnamefont{Drechsel}},
  \bibinfo{author}{\bibfnamefont{A.}~\bibnamefont{Metz}}, \bibnamefont{and}
  \bibinfo{author}{\bibfnamefont{M.}~\bibnamefont{Vanderhaeghen}},
  \bibinfo{journal}{Eur. Phys. J.} \textbf{\bibinfo{volume}{A11}},
  \bibinfo{pages}{185} (\bibinfo{year}{2001}), \eprint{hep-ph/0102335}.

\bibitem[{\citenamefont{Drechsel et~al.}(2003)\citenamefont{Drechsel, Pasquini,
  and Vanderhaeghen}}]{Drechsel:2002ar}
\bibinfo{author}{\bibfnamefont{D.}~\bibnamefont{Drechsel}},
  \bibinfo{author}{\bibfnamefont{B.}~\bibnamefont{Pasquini}}, \bibnamefont{and}
  \bibinfo{author}{\bibfnamefont{M.}~\bibnamefont{Vanderhaeghen}},
  \bibinfo{journal}{Phys. Rept.} \textbf{\bibinfo{volume}{378}},
  \bibinfo{pages}{99} (\bibinfo{year}{2003}), \eprint{hep-ph/0212124}.

\bibitem[{\citenamefont{Blomqvist et~al.}(1998)}]{Blomqvist:1998xn}
\bibinfo{author}{\bibfnamefont{K.~I.} \bibnamefont{Blomqvist}}
  \bibnamefont{et~al.}, \bibinfo{journal}{Nucl. Instrum. Meth.}
  \textbf{\bibinfo{volume}{A403}}, \bibinfo{pages}{263} (\bibinfo{year}{1998}).

\bibitem[{\citenamefont{Beri\v{c}i\v{c}}(2015)}]{BericicPhD:2015}
\bibinfo{author}{\bibfnamefont{J.}~\bibnamefont{Beri\v{c}i\v{c}}}, Ph.D.
  thesis, \bibinfo{school}{{University of Ljubljana}} (\bibinfo{year}{2015}).

\bibitem[{\citenamefont{Correa}(2016)}]{CorreaPhD:2016}
\bibinfo{author}{\bibfnamefont{L.}~\bibnamefont{Correa}}, Ph.D. thesis,
  \bibinfo{school}{{Clermont-Fd and Mainz Universities}}
  (\bibinfo{year}{2016}).

\bibitem[{\citenamefont{BenAli}(2016)}]{BenaliPhD:2016}
\bibinfo{author}{\bibfnamefont{M.}~\bibnamefont{BenAli}}, Ph.D. thesis,
  \bibinfo{school}{{Clermont-Fd University}} (\bibinfo{year}{2016}).

\bibitem[{\citenamefont{Janssens et~al.}(2006)}]{Janssens:2006vx}
\bibinfo{author}{\bibfnamefont{P.}~\bibnamefont{Janssens}}
  \bibnamefont{et~al.}, \bibinfo{journal}{Nucl. Instrum. Meth.}
  \textbf{\bibinfo{volume}{A566}}, \bibinfo{pages}{675} (\bibinfo{year}{2006}),
  \eprint{physics/0608308}.

\bibitem[{\citenamefont{Friedrich and Walcher}(2003)}]{Friedrich:2003iz}
\bibinfo{author}{\bibfnamefont{J.}~\bibnamefont{Friedrich}} \bibnamefont{and}
  \bibinfo{author}{\bibfnamefont{T.}~\bibnamefont{Walcher}},
  \bibinfo{journal}{Eur. Phys. J.} \textbf{\bibinfo{volume}{A17}},
  \bibinfo{pages}{607} (\bibinfo{year}{2003}), \eprint{hep-ph/0303054}.

\bibitem[{\citenamefont{Guichon and Vanderhaeghen}(1998)}]{Guichon:1998xv}
\bibinfo{author}{\bibfnamefont{P.~A.~M.} \bibnamefont{Guichon}}
  \bibnamefont{and}
  \bibinfo{author}{\bibfnamefont{M.}~\bibnamefont{Vanderhaeghen}},
  \bibinfo{journal}{Prog. Part. Nucl. Phys.} \textbf{\bibinfo{volume}{41}},
  \bibinfo{pages}{125} (\bibinfo{year}{1998}), \eprint{hep-ph/9806305}.

\bibitem[{\citenamefont{d'Hose}(2006)}]{dHose:2006bos}
\bibinfo{author}{\bibfnamefont{N.}~\bibnamefont{d'Hose}},
  \bibinfo{journal}{Eur. Phys. J.} \textbf{\bibinfo{volume}{A28S1}},
  \bibinfo{pages}{117} (\bibinfo{year}{2006}).

\bibitem[{\citenamefont{N.Sparveris et~al.}(2016)}]{Sparveris:2016}
\bibinfo{author}{\bibnamefont{N.Sparveris}} \bibnamefont{et~al.}
  (\bibinfo{year}{2016}), \bibinfo{note}{{J}Lab Proposal PR12-15-001}.

\bibitem[{\citenamefont{Pasquini et~al.}(2019)\citenamefont{Pasquini, Pedroni,
  and Sconfietti}}]{Pasquini:2019nnx}
\bibinfo{author}{\bibfnamefont{B.}~\bibnamefont{Pasquini}},
  \bibinfo{author}{\bibfnamefont{P.}~\bibnamefont{Pedroni}}, \bibnamefont{and}
  \bibinfo{author}{\bibfnamefont{S.}~\bibnamefont{Sconfietti}}
  (\bibinfo{year}{2019}), \eprint{1903.07952}.

\bibitem[{\citenamefont{Lensky et~al.}(2017)\citenamefont{Lensky, Pascalutsa,
  and Vanderhaeghen}}]{Lensky:2016nui}
\bibinfo{author}{\bibfnamefont{V.}~\bibnamefont{Lensky}},
  \bibinfo{author}{\bibfnamefont{V.}~\bibnamefont{Pascalutsa}},
  \bibnamefont{and}
  \bibinfo{author}{\bibfnamefont{M.}~\bibnamefont{Vanderhaeghen}},
  \bibinfo{journal}{Eur. Phys. J.} \textbf{\bibinfo{volume}{C77}},
  \bibinfo{pages}{119} (\bibinfo{year}{2017}), \eprint{1612.08626}.

\bibitem[{\citenamefont{Tanabashi et~al.}(2018)}]{Tanabashi:2018oca}
\bibinfo{author}{\bibfnamefont{M.}~\bibnamefont{Tanabashi}}
  \bibnamefont{et~al.} (\bibinfo{collaboration}{Particle Data Group}),
  \bibinfo{journal}{Phys. Rev.} \textbf{\bibinfo{volume}{D98}},
  \bibinfo{pages}{030001} (\bibinfo{year}{2018}).

\end{thebibliography}
\end{document}